\documentclass[final,3p,times,twocolumn]{elsarticle}
 \biboptions{comma,sort&compress}
\usepackage{here}
 \usepackage{graphicx}
  \usepackage{epsfig}
\def\beq{\begin{equation}}
\def\eeq{\end{equation}}
\def\bea{\begin{eqnarray}}
\def\eea{\end{eqnarray}}

\def\z0{\rm Z^0}
\newcommand{\as}{\alpha_{\rm s}}

\newcommand{\oaaa}{{\cal O}(\as^3)}

\newcommand{\amz}{\as(M_{\rm Z})}

\def\d2{D_2}
\def\oq{\char'134}
\def\lamsb{\Lambda_{\overline{MS}}}

\def\m2{\mu^2}
\def\q{\rm q}

\def\x{\rm X}
\def\q2{Q^2}

\def\asq{\as (\q2 )}

\journal{Nuc. Phys. (Proc. Suppl.)}

\begin{document}

\begin{frontmatter}



\title{World Summary of $\alpha_s$ (2012)\tnoteref{label1}}
\tnotetext[label1]{Talk presented at 
$16^{th}$ International Conference 
in Quantum ChromoDynamics (QCD 12)
$2-6^{th}$ July 2012 (Montpellier - France) }

\author{Siegfried Bethke}
\address{Max-Planck-Institute for Physics, F\"ohringer Ring 6, 80805 Munich, Germany}
\ead{bethke@mpp.mpg.de}

\begin{abstract}
\noindent
Determinations of the strong coupling strength, $\alpha_s$,
are summarised and a new world average value of $\alpha_s (M_Z)$ is determined,
using a new method of pre-averaging results within classes of measurements like
hadronic $\tau$ decays, deep inelastic scattering processes,
lattice calculations, electron-positron annihilation processes and electro-weak precision fits.
The overall result is
$$\alpha_s (M_Z) = 0.1184 \pm 0.0007\ ,$$
unchanged from the value obtained in 2009.
This presentation is an excerpt from the QCD review section of the 2012
Review of Particle Physics (RPP) of the Particle Data Group \cite{pdg2012}.
An earlier version of this work was also given in \cite{ringberg11}.

\end{abstract}

\begin{keyword}
strong interactions \sep QCD \sep strong coupling
\end{keyword}

\end{frontmatter}

% \linenumbers


\section{Introduction}
\label{intro}

The size of the strong coupling strength, $\as$, like other
fundamental \oq constants" of nature, is not given by theory, but must be determined 
by experiment.
Quantum Chromodynamics (QCD) \cite{qcd}, the gauge field theory describing 
the Strong Interaction between quark and gluons, the fundamental constituents of 
hadronic matter, allows to predict physical cross sections and many other
observables $\cal R$ of particle reactions involving quarks and gluons, in terms of - basically -
one single parameter, $\as$.
Assuming $\as < 1$ and applying perturbation theory, predictions are typically given by a
power series in $\as$, like:
\begin{eqnarray} \label{eq-rseries}
{\cal R} &=&   P_{l} \sum_{n} R_n \as^n \nonumber \\              
&=& P_l \left( R_0 + R_1 \as  + R_2  \as^2 + ...\right)
\end{eqnarray}\noindent 
where $R_n$ are the $n_{th}$ order coefficients of the perturbation series 
and $P_l R_0$ denotes the lowest-order value of $\cal R$.
This allows to measure $\as$ in a large variety of particle reactions.

QCD also predicts the energy dependence of $\as$ through the
\oq beta function",
\begin{equation} \label{eq-rge}
Q^2 \frac{\partial \asq}{\partial Q^2} = \beta \left( \asq \right) \ ,
\end{equation} 
\noindent
where $\q2$ is the squared energy scale or the momentum
transferred of the particle reaction under study,
and any energy scale dependence of $\cal R$ is determined by the energy dependence of $\as$\footnote{
For processes with initial state hadrons, also the coefficients $R_n$ may
depend on $\as$, through the effects of parton density functions.}.
Measurements of $\as$ from different particle reactions and scattering processes,
performed at different energy scales $Q^2$, therefore $test$ the global nature of QCD and, in particular,
its characteristic prediction of \oq Asymptotic Freedom", which determines that $\as$ is
small and asymptotically tends to zero at large energy scales (or at small distances), while it 
is large at small energy scales (large distances), explaining the \oq confinement" of quarks and gluons
inside hadrons.

Experimental determinations of $\as$ were regularly summarised and reviewed
in the past, see e.g. \cite{as2006,as2009,pdg2010}.
These references also contain the definition of basic equations and fomulae which
shall not be repeated here.
In 2009, the world average value of $\as$, expressed at a common energy scale
corresponding to the rest mass of the $Z$ boson, determined from a set of
most recent determinations from many different processes at a large range of energy
scales, converged to
$$\amz = 0.1184 \pm 0.0007\ ,$$
where the overall error includes experimental as well as (dominating) systematic
and theoretical uncertainies \cite{as2009}.
In the following, this summary is updated with the newest results in this field.

\section{Selected results}

The set of results used in this review is restricted to those
which were published in peer-reviewed journals until April 2012, and
which use QCD perturbation theory to at least full next-to-next-to-leading order 
perturbation\footnote{NNLO; for observables with QCD contributions starting at leading order, 
that is $\oaaa$.} .

These requirements exclude e.g. results from jet production in deep inelastic lepton-nucleon
scattering (DIS) and from hadron collisions,
as well as those from heavy quarkonia decays for which calculations are available in NLO only.
While being excluded from calculating a world average value of $\as$,
NLO results will nevertheless be cited in this review as they are important
ingredients for demonstrating 
the experimental evidence of the energy dependence of $\alpha_s$, i.e. for 
Asymptotic Freedom, one of the key features of QCD.

Furthermore, 
here we add an intermediate step of pre-averaging results 
within well defined sub-fields like $e^+e^-$-annihilation, 
DIS and hadronic $\tau$-decays. 
The overall world average is then  calculated from those
pre-averages rather than from individual measurements.
The intermediate step is done because in a number of sub-fields, different
determinations of the strong coupling from substantially similar datasets 
lead to values of $\alpha_s$ that are only marginally compatible with each other,
or with the final world average value, which may be a reflection
of the challenges of evaluating systematic uncertainties.
In such cases, a pre-average value is determined, with a symmetric, overall error that encompasses
the central values of all individual determinations.

\subsection{Hadronic $\tau$ decays} 
Determination of $\as$ from hadronic $\tau$ lepton decays 
continues to be one of the most actively studied fields to measure this basic
quantity.
The small effective energy scale, $Q = M_\tau =
1.78$~GeV, small nonperturbative contributions to 
an inclusive and well defined experimental observable,
and the availability of perturbative predictions which are complete to N$^3$LO 
determine the importance and large interest in this particular field.
Several re-analyses of the hadronic $\tau$ decay 
width \cite{Baikov:2008jh,Beneke:2008ad,Davier:2008sk,Maltman:2008nf,Narison:2009vy, Caprini:2009vf}
were performed, using different approaches of treating perturbative
(fixed order or contour improved perturbative expansions) and non-perturbative contributions. 
The result of \cite{Baikov:2008jh} 
includes both, fixed order and 
contour improved perturbation, while the others adhere to either one or the other of the two.
These results are summarised in Fig.~\ref{fig:pre-averages}(a).

There are more studies of $\alpha_s$ from $\tau$-decays,
\cite{Menke:2009vg,Pich:2011xy,Magradze,Abbas}, which were not yet available as
peer-reviewd publications.
They are compatible with the overall picture as 
summarized in Fig.~\ref{fig:pre-averages}(a).
Another study \cite{Boito:2011} argues that an improved treatment of
non-perturbative effects results in  values of $\alpha_s$ which
are systematically lower than those discussed above.
These results have recently been extended in \cite{Boito:2012}.

The pre-average result from $\tau$-decays, to be used
for calculating the final world average of $\alpha_s(M_Z^2)$, 
is determined using the simple
method mentioned above, 
i.e. defining one central value with symmetric overall error bars which include the
smallest as well as the largest of all results, as 
$\alpha_s(M_{\tau}^2) = 0.330 \pm 0.014$.
This value of $\alpha_s(M_{\tau}^2)$ corresponds, when evolved to the scale of the Z-boson, using
the QCD 4-loop beta-function plus 3-loop matching at the charm- and the bottom-quark
masses (see \cite{as2009,as2006,pdg2010} for relevant equations), 
to $\alpha_s(M_Z^2) = 0.1197 \pm 0.0016$,
unchanged from its value in the 2009 review.

\subsection{Lattice QCD} 
There are several recent results on $\alpha_s$ from lattice QCD, see
also Sec. {\it Lattice QCD} in \cite{pdg2012}.
The HPQCD collaboration \cite{Davies:2010sw} computes
     short distance quantities like
     Wilson loops with lattice QCD and analyzes them with NNLO
     perturbative QCD.  This yields a value for
     $\alpha_s$. 
     The lattice scale must then be related to a physical energy/momentum scale.  
     This is achieved with the $\Upsilon'$-$\Upsilon$ mass difference, however, many other quantities could
   be used as well \cite{Davies:2003ik}.  HPQCD obtains $\alpha_s(M_Z^2) =  0.1184 \pm 0.0006$, 
   where the uncertainty includes effects from truncating perturbation theory, finite lattice spacing
   and extrapolation of lattice data.  An independent perturbative analysis of a subset of the same lattice-QCD data 
   yields $\alpha_s(M_Z^2)=0.1192 \pm 0.0011$\cite{Maltman:2008bx}.  
Using another, independent methodology, the current-current correlator method, HPQCD obtains
$\alpha_s(M_Z^2) =  0.1183 \pm 0.0007$ \cite{Davies:2010sw}.
   A more recent result in \cite{Aoki:2009tf}, which avoids
   the staggered fermion treatment of \cite{Davies:2010sw}, finds
   $\alpha_s(M_Z^2) = 0.1205 \pm 0.0008 \pm 0.0005\,^{+0.0000}_{-0.0017}$\cite{Aoki:2009tf},
   where the first uncertainty is statistical and the others are from systematics. 
   Since this approach uses a different discretization of lattice fermions 
   and a different general methodology, it provides an independent cross check
   of other lattice extractions of $\alpha_s$.
Finally, the JLQCD collaboration - in an analysis of Adler functions - obtains 
$\alpha_s(M_Z^2) = 0.1181 \pm 0.0003\,^{+0.0014}_{-0.0012}$\cite{JLQCD:2010xy}.

These results are summarized in Fig.~\ref{fig:pre-averages}(b).
Since they are compatible with and largely independent from each other, a pre-average
of lattice results is calculated
using the same method as applied to determine the
final world average value $\as$, i.e. calculate a weighted
average and a (correlated) error such that the overall $\chi^2$ equals unity per
degree of freedom - rather than using the simple method as applied in the case of
$\tau$ decays.  
This gives 
$\amz = 0.1185 \pm 0.0007$
which is taken as result from the sub-field of lattice determinations.

{\subsection{Deep inelastic lepton-nucleon scattering (DIS)} 
Studies of DIS final states have
led to a number of precise determinations of $as$: \\
A combination \cite{Glasman:2007sm} 
of precision measurements at HERA, based on NLO fits to inclusive jet cross sections
in neutral current DIS at high $Q^2$, 
quotes a combined result of   $\amz =  0.1198 \pm 0.0032$, which
includes a theoretical uncertainty of $\pm 0.0026$.
 A combined analysis of non-singlet structure functions from DIS \cite{Blumlein:2006be}, 
based on QCD predictions up to 
N$^3$LO in some of its parts, 
gave $\amz =  0.1142 \pm 0.0023$,
including a theoretical error of $\pm 0.0008$.
Further studies of singlet and non-singlet structure functions, based on NNLO predictions,
resulted in $\amz =  0.1129 \pm 0.0014$
\cite{abkm:2010ab} and in 
$\amz =  0.1158 \pm 0.0035$
\cite{jr:2009ab}.
The MSTW group \cite{mstw:2009ab}, also including data on jet
production at the Tevatron, obtains, in NNLO\footnote{Note that
for jet production at the hadron collider, only NLO predictions are available,
while for the structure functions full NNLO was utilized.},
$\amz =  0.1171 \pm 0.0014$.
Most recently, the NNPDF group \cite{nnpdf} has presented a result,
$\amz = 0.1173 \pm 0.0011$, which is in line with the one from the MSTW group.

Summarizing these results from world data on structure functions,
applying the same method as in the case of summarizing results from
$\tau$ decays, leads
to a pre-average value of $\amz =  0.1151 \pm 0.0022$
(see Fig.~\ref{fig:pre-averages}(c)).

Note that criticism has been expressed on some of the above extractions.
Among the issues raised, we mention the neglect of singlet contributions
at $x \ge 0.3$ in pure non-singlet fits \cite{thorne:2011}, the impact
and detailed treatment of particular classes of data in
the fits \cite{1101.5261,1102.3182,thorne:2011} and possible biases due to
insufficiently flexible parametrizations of the PDFs \cite{1102.3182}.

\subsection{Heavy quarkonia decays}
The most recent extraction of the strong coupling 
constant from an analysis of radiative $\Upsilon$ 
decays \cite{Brambilla:2007cz} resulted in $\amz =  0.119^{+0.006}_{-0.005}$.
This determination is based on QCD in NLO only, so it will not be considered for the
final extraction of the world average value of $\alpha_s$; it is, however, an important ingredient
for the demonstration of Asymptotic Freedom as given in Fig.~\ref{fig:runningas}.

\subsection{Hadronic final states of $e^+e^-$ annihilations} 
Re-analyses of event shapes in $e^+e^-$-annihilation, measured at the $Z$ peak
and LEP2 energies up to 209 GeV, using NNLO predictions matched to NLL resummation, resulted in
$\amz =  0.1224\pm0.0039$ \cite{Dissertori:2009ik}, with a dominant theoretical uncertainty of $0.0035$,
and in $\amz =  0.1189\pm0.0043$ \cite{opal:2011ab}.
Similarly, an analysis of JADE data \cite{Bethke:2008hf} at 
center-of-mass energies between 14 and 46 GeV gives
$\amz =  0.1172\pm0.0051$, with contributions from hadronization model (perturbative QCD)
uncertainties of $0.0035\, (0.0030)$.
A precise determination of $\alpha_s$ from 3-jet production alone, in NNLO,
resulted in
$\amz =  0.1175\pm0.0025$ \cite{dissertori:2010ab}.
Computation of the NLO corrections to 5-jet production and comparison
to the measured 5-jet rates at LEP \cite{frederix:2010ab} 
gave 
$\amz =  0.1156^{+0.0041}_{-0.0034}$.
More recently, a study using the world data of Thrust distributions and soft-collinear effective theory,
including fixed order NNLO, gave
$\amz =  0.1135\pm0.0010$ \cite{hoang:2010ab}.

Note that there is criticism on both classes of $\alpha_s$ extractions just described:
those based on corrections of non-perturbative hadronisation effects using QCD-inspired
Monte Carlo generators (since the parton level of a Monte Carlo is not defined in a manner equivalent to that of a fixed-order calculation),
as well as the studies based on effective field theory, as their systematics
have not yet been verified e.g. by using observables other than Thrust.

A summary of the $e^+e^-$ results based on NNLO predictions is shown in 
Fig.~\ref{fig:pre-averages}(d).
They average, according to the simple procedure defined above, to
$\amz =  0.1172\pm0.0037$.

\begin{figure*}[hbt]
%\resizebox{0.49\textwidth}{!}{%
  \includegraphics{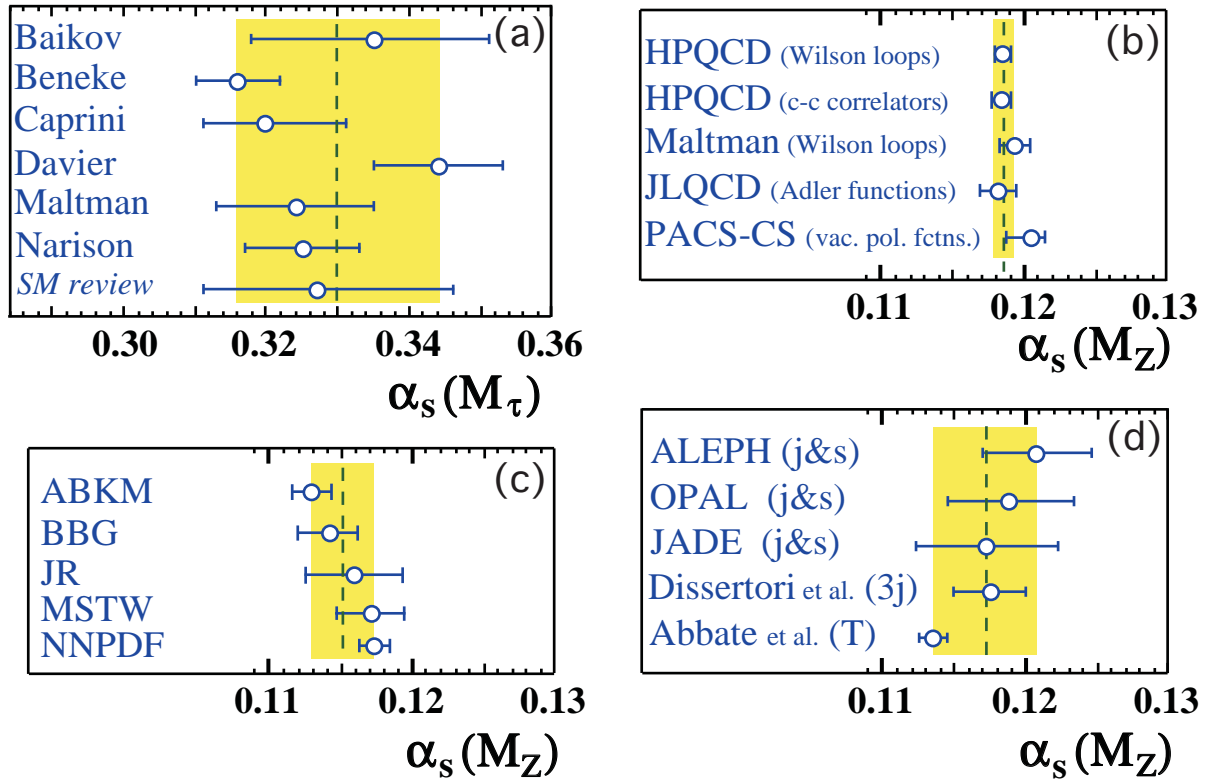}
\caption{Summary of determinations of $\alpha_s$ from 
hadronic $\tau$-decays (a), from lattice calculations (b),
from DIS structure functions (c) and from
event shapes and jet production in
$e^+e^-$-annihilation (d). 
The shaded bands indicate the average values chosen 
to be included in the determination of the new world average of
$\alpha_s$.
Figure taken from \cite{pdg2012}.}
\label{fig:pre-averages}       
\end{figure*}

\subsection{Hadron collider jets} 
A determination of $\alpha_s$ from
the $p_T$ dependence
of the inclusive jet cross section in $p\overline{p}$ collisions at $\sqrt{s}$ = 1.96 TeV, in
the transverse momentum range  of $50 < p_T < 145$~GeV, based on NLO 
($\cal O \rm (\alpha_s^3)$) QCD, led to
$\amz =  0.1161^{+0.0041}_{-0.0048}$ \cite{abazov:2009ab}, 
which is the most precise $\alpha_s$ result obtained
at a hadron collider.
Experimental uncertainties from the jet
energy calibration, the $p_T$ resolution and the integrated
luminosity dominate the overall error.

\subsection{Electroweak precision fits} 
The N$^3$LO calculation of the hadronic $Z$ decay width 
was used in a recent revision of the global
fit to electroweak precision data \cite{flacher:2009a}, resulting in  $\amz =  0.1193 \pm 0.0028$,
claiming a negligible theoretical uncertainty.
For this review, we use the result 
obtained in \cite{pdg2012}
from data at the $Z$-pole, 
$\alpha_s(M_Z^2) =  0.1197 \pm 0.0028$,
as it is based on a more constrained data set
where QCD corrections directly enter through the hadronic decay width of the $Z$.
Note that all these results 
from electroweak precision data, however, strongly depend on the strict validity of Standard Model predictions
and the existence of the minimal Higgs mechanism to implement electroweak symmetry breaking.
Any - even small - deviation of nature from this model could strongly influence this extraction of $\alpha_s$.

\section{Determination of the world average value of $\amz$} 
A non-trivial exercise consists in the evaluation of a world-average 
value for $\amz$. A certain arbitrariness 
and subjective component is inevitable because of the choice of measurements to be included in the average,
the treatment of (non-Gaussian) systematic uncertainties of mostly theoretical nature, as well as the
treatment of correlations among the various inputs,  of theoretical as well as experimental origin. In 
earlier reviews \cite{as2009,pdg2010,as2006} 
an attempt was made to take account of such correlations, 
using methods as proposed e.g. in \cite{Schmelling:1994pz}, and - likewise - to treat 
cases of apparent incompatibilities or possibly underestimated systematic uncertainties
in a meaningful and well defined manner: 

The central value is determined
as the weighted average of the different input values. 
An initial  error of the central value is determined treating
the uncertainties of all individual measurements as being uncorrelated and being
of Gaussian nature, and the overall $\chi^2$ to the central value is determined.
In case this initial $\chi^2$ is larger than the number of degrees of freedom,
i.e. larger than the number of individual inputs minus one, then all individual
errors are enlarged by a common factor such that $\chi^2 / d.o.f.$ equals unity.
If the initial value of $\chi^2$ is smaller than the number of degrees of freedom,
an overall, a-priori unknown correlation coefficient
is introduced and determined by requiring that the total $\chi^2 / d.o.f.$
of the combination equals unity.
In both cases, the resulting final overall uncertainty of the central value
of $\alpha_s$ is larger than the initial estimate of a Gaussian error.

This procedure is only meaningful if the individual measurements are known not to be 
correlated to large degrees, i.e. if they are not - for instance - based on
the same input data, and if the input values are largely compatible with each other
and with the resulting central value, within their assigned uncertainties.
The list of selected individual measurements discussed above, however, 
violates both these requirements:
there are several measurements based on (partly or fully) identical data sets,
and there are results which apparently do not agree with others and/or with the resulting
central value, within their assigned individual uncertainty.
Examples for the first case are results from the hadronic width of the $\tau$ lepton,
from DIS processes and from jets and event shapes in $e^+e^-$ final states.
An example of the second case is the apparent disagreement between results 
from the $\tau$ width and those from DIS 
\cite{Blumlein:2006be} or from Thrust distributions in $e^+e^-$
annihilation \cite{hoang:2010ab}.

Due to these obstacles, we have chosen to determine pre-averages for each class of
measurements, and then to combine those to the final world average
value of $\amz$, using the methods of error treatment as just described.
The five pre-averages are summarized in Fig.~\ref{fig:as-summary};
we recall that these are exclusively obtained from extractions which are based on
(at least) full NNLO QCD predictions, and are published in peer-reviewed journals 
at the time of completing this review.
From these, the new central and world average value of
\begin{equation}
    \amz =  0.1184 \pm 0.0007  \ ,
\end{equation}
\noindent
is determined,
with an uncertainty of well below 1~\%.\footnote{The weighted average, 
treating all inputs as uncorrelated measurements with Gaussian errors, results in
$\amz =  0.11844 \pm 0.00059$ with $\chi^2$/d.o.f. = 3.2/4.
Requiring $\chi^2$/d.o.f. to reach unity leads to a common correlation factor
of 0.19 which increases the overall error to 0.00072.}
This world average value is - in spite of several new contributions to this
determination - identical to and thus, in excellent agreement with
the 2009 result \cite{as2009,pdg2010}.
For convenience, we also provide corresponding values for $ \lamsb$
suitable for use with the common $\Lambda$-parametrisation
of $\as$, see e.g. Eq. 6 in \cite{as2009}:
\begin{eqnarray}
\Lambda^{(5)}_{\overline{MS}} &=& (213 \pm 8)~{\rm MeV}\ ,\\
\Lambda^{(4)}_{\overline{MS}} &=& (296 \pm 10)~{\rm MeV}\ ,\\
\Lambda^{(3)}_{\overline{MS}} &=& (339 \pm 10)~{\rm MeV}\ ,
\end{eqnarray}
\noindent for $N_f = 5$, 4 and 3 quark flavors, respectively.

\begin{figure}
\resizebox{0.49\textwidth}{!}{%
  \includegraphics{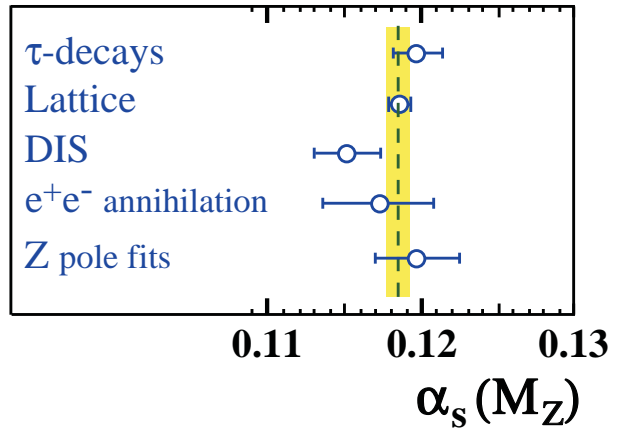}}
\caption{Summary of values of $\amz)$ obtained for
various sub-classes of measurements (see Fig.~\ref{fig:pre-averages} (a) to (d)).
The new central and world average value of 
$\amz = 0.1184 \pm 0.0007$ is indicated by the
dashed line and the shaded band.
Figure taken from \cite{pdg2012}.}
\label{fig:as-summary}       
\end{figure}

In order to verify the consistency and stability of the new result, we give each of the averages
obtained when leaving out one of the five input values:
\begin{eqnarray}
\amz &=&  0.1182 \pm 0.0007\ \ {\rm(w/o\ \tau\ results)}, \nonumber\\
\amz &=&  0.1183 \pm 0.0012\ \ {\rm(w/o\ lattice)}, \nonumber\\
\amz &=&  0.1187 \pm 0.0009\ \ {\rm(w/o\ DIS)}, \nonumber\\
\amz &=&  0.1185 \pm 0.0006\ \ {\rm(w/o\ e^+e^-),\ and} \nonumber\\
\amz &=&  0.1184 \pm 0.0006\ \ {\rm(w/o\ e.w.\ prec.\ fit)}.\ \nonumber
\end{eqnarray}
They are well within the error of the overall world average quoted above.
Most notably, the result from lattice calculations, which has the
smallest assigned error, agrees well with the exclusive average
of the other results. However, it largely determines the size of the (small) overall
uncertainty.

There are apparent systematic differences between the
various structure function results, and also between the new result from Thrust in
$e^+e^-$ annihilation and the other determinations.
Expressing this in terms of a $\chi^2$ between a given measurement and the world
average as obtained when {\it excluding} that particular measurement, the largest values are $\chi^2 = 12.6$ 
and $\chi^2 = 16.1$, corresponding to 3.5 and 4.0 standard deviations,
for the measurements of \cite{abkm:2010ab} and \cite{hoang:2010ab}, respectively.
We note that such and other differences have been extensively discussed
at a recent workshop on measurements of $\alpha_s$, however 
none of the explanations proposed so far have obtained enough of a consensus to definitely
resolve these tensions \cite{alphasworkshop:2011}.

Notwithstanding these open issues, a rather stable and well defined
world average value emerges from the compilation of current
determinations of $\alpha_s$:
$$
    \amz =  0.1184 \pm 0.0007  \ .
$$
\noindent
The results also provide a clear signature and proof of the
energy dependence of $\alpha_s$, in full agreement with the QCD
prediction of Asymptotic Freedom.  This is demonstrated in Fig.~\ref{fig:runningas},
where results of $\alpha_s(Q^2)$ obtained at discrete energy scales
$Q$, now also including those based just on NLO QCD, are summarized.

\begin{figure}
\resizebox{0.49\textwidth}{!}{%
  \includegraphics{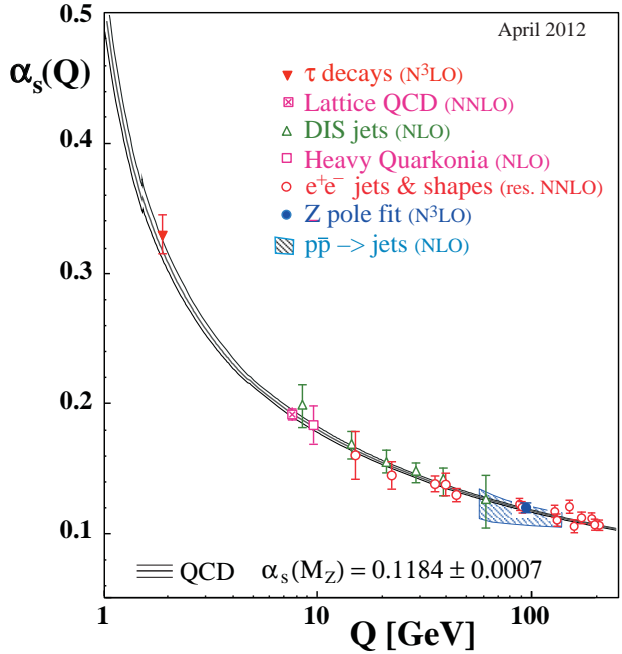}}
\caption{Summary of measurements of $\alpha_s$ as a function of the respective energy scale $Q$.
The respective degree of QCD perturbation theory used in 
the extraction of $\alpha_s$ is indicated in brackets (NLO: next-to-leading order;
NNLO: next-to-next-to leading order; res. NNLO: NNLO matched with resummed
next-to-leading logs; N$^3$LO: next-to-NNLO).
Figure taken from \cite{pdg2012}.}
\label{fig:runningas}       
\end{figure}

\vspace{1cm}
\noindent \textbf{Acknowledgments:}
Many thanks go to Stephan Narison and his team for organising this pleasant and interesting workshop.
I am grateful to G. Dissertori and G. Salam, my co-authors for the QCD section of the
2012 edition of the Review of Particle Physics, for their valuable collaboration.

% ----------------------------------------------------------------------------

\end{document}